\documentclass[useAMS,usenatbib]{mn2e}

\title[Kinematics of NGC5694]
  {{Kinematics of a globular cluster with an extended profile: NGC~5694.}
  \thanks{Based on data obtained at the Very Large Telescope under the program 089.D-0094.}}
\author[M. Bellazzini et al.]{M. Bellazzini$^{1}$\thanks{E-mail:
michele.bellazzini@oabo.inaf.it}, A. Mucciarelli$^2$, A. Sollima$^{1}$, M. Catelan$^{3,9}$,
E. Dalessandro$^{2}$, 
\newauthor
M. Correnti$^{4}$, , V. D'Orazi$^{5,6}$, C. Cort\'es$^{7,9}$, P. Amigo$^{8,9}$\\
$^{1}$INAF-Osservatorio Astronomico di Bologna, Via Ranzani 1, 40127, Bologna, Italy.\\
$^{2}$Dip. di Astronomia - Univ. di Bologna, Via Ranzani 1, 40127, Bologna, Italy.\\
$^{3}$ Instituto de Astrof\'isica, Pontificia Universidad Cat\'olica de Chile, Av. Vicuna Mackenna 4860,\\ 
782-0436 Macul, Santiago, Chile\\
$^{4}$ Space Telescope Science Institute, 3700 San Martin Drive, Baltimore, MD 21218, USA\\
$^{5}$Department of Physics and Astronomy, Macquarie University, Sydney, NSW 2109, Australia\\ 
$^{6}$Monash Centre for Astrophysics, School of Mathematical Sciences, Monash University, Melbourne, VIC 3800, Australia\\
$^{7}$ Departamento de F\'isica, Facultad de Ciencias B\'asicas, Universidad Metropolitana de Ciencias de la Educaci\'on,\\ 
$^{8}$ Departamento de F\'isica y Astronom\'ia, Universidad de Valpara\'iso, Av. Gran Breta\~na 1111, Playa Ancha, Valpara\'iso, Chile\\
Av. Jos\'e Pedro Alessandri 774, 776-0197 \~Nu\~noa, Santiago, Chile\\
$^{9}$ Millennium Institute of Astrophysics, Santiago, Chile
}

\usepackage{graphicx}

\pagerange{\pageref{firstpage}--\pageref{lastpage}} \pubyear{2014}

\def\LaTeX{L\kern-.36em\raise.3ex\hbox{a}\kern-.15em
    T\kern-.1667em\lower.7ex\hbox{E}\kern-.125emX}

\begin{document}
\date{Accepted for publication by MNRAS on October 29, 2014}

\label{firstpage}

\maketitle

\begin{abstract}

We present a study of the kinematics of the remote globular cluster NGC~5694 based on GIRAFFE@VLT medium resolution spectra. A sample of 165 individual stars selected to lie on the Red Giant Branch in the cluster Color Magnitude Diagram was considered. Using radial velocity and metallicity from Calcium triplet, we were able to select 83 bona-fide cluster members. The addition of six previously known members leads to a total sample of 89 cluster giants with typical uncertainties $\le 1.0$~km/s in their radial velocity estimates. The sample covers a wide range of projected distances from the cluster center, from $\sim 0.2\arcmin$ to 
$6.5\arcmin \simeq 23$ half-light radii ($r_h$). We find only very weak rotation, as typical of metal-poor globular clusters. The velocity dispersion gently declines from a central value of $\sigma=6.1$~km/s to $\sigma\simeq 2.5$~km/s at $\sim 2\arcmin \simeq 7.1 r_h$, then it remainins flat out to the next (and last) measured point of the dispersion profile, at $\sim 4\arcmin \simeq 14.0 r_h$, at odds with the predictions of isotropic King models.
We show that both isotropic single-mass non-collisional models and multi-mass anisotropic models can reproduce the observed surface brightness and velocity dispersion profiles.

\end{abstract}

\begin{keywords}
(Galaxy:) globular clusters: individual: NGC 5694 -- stars: abundances
\end{keywords}

\section{Introduction}

NGC~5694 is a bright ($M_V=-8.0$) and remote ($D=35.5$~kpc) old and metal-poor Galactic
Globular Cluster (GC), located in the Hydra constellation. 
First discovered by W. Herschel in 1784, it has been
recognised as a GC by \citet{latom}. Because of its distance
and low apparent magnitude, the first photometric studies of giant stars
in this cluster have been conducted only in relatively recent epoch \citep{H75,orto}. 
After the first integrated spectroscopic studies
\citep[see, e.g.,][]{HH76} spectroscopy of individual Red Giant stars in NGC~5694
have been carried out by \citet{geisler95} and, more recently, by \citet{lee}.
The latter derived the chemical composition for one bright giant of the cluster
from a high-resolution spectrum and found an abundance pattern different from ordinary
stars and clusters in the Galactic Halo, more similar to those displayed by stars in dwarf spheroidal galaxies.
The accretion of globular clusters into the halo of giant galaxies during the disruption of their parent dwarf galaxy is now established to have occurred in the Milky Way \citep[see, e.g.,][and references therein]{b03,lw10,carba} and in M31 \citep{peri,mack10,mack13}. Dense nuclei of stripped dwarf satellites can also appear as massive GCs at the present epoch \cite[see][for references and discussion]{bekki,b08,seth}.
Chemical tagging is one of the main technique to identify otherwise ordinary GCs as accreted from a former, and now fully disrupted, Galactic satellite.

To follow-up the intriguing finding by \citet{lee} we started a multi-instrument observational campaign that allowed us (a) to trace the surface brightness (SB)
profile of the cluster down to $\mu_V\simeq 30.0$~mag/arcsec$^2$, finding that it extends smoothly much 
beyond the tidal radius of the best-fitting \citet{king66} model and that it cannot be adequately fit 
neither by a \citet{king66}, \citet{wilson} nor \citet{elson} model \citep[][C11 hereafter]{correnti}, and (b) to 
perform accurate abundance analysis from high-resolution spectra for {\em six} cluster giants, fully 
confirming that the cluster has a chemical pattern different from the Galactic Halo, with nearly solar 
[$\alpha$/Fe] ratio and anomalously low abundances of Y, Ba, La and Eu, at [Fe/H]$\simeq -2.0$ \citep[][Mu13 hereafter]{m13_5694}.

Here we present a study of the kinematics of the cluster, based on a large sample of medium-resolution
spectra of stars selected to lie on the Red Giant Branch (RGB) in the color magnitude diagram of the cluster. 
The plan of the paper is the following: in Sect.~2 we present our observations and we describe the data reduction. 
In Sect.~3 and 4 we describe how we derived our estimates of the radial velocity and metallicity, respectively, 
from the available spectra. In Sect.~5 we present our criteria to select cluster members and the analysis of the cluster 
kinematics, including estimates of the dynamical mass. Finally in 
Sect.~6 we briefly summarise and discuss the results of the analysis.

\section{Observations}

%%%%%%OSSERVAZIONI
The data have been acquired with the multi-object facility FLAMES@VLT \citep{pasquini} 
in the combined MEDUSA+UVES mode, allowing the simultaneous allocation of 8 UVES high-resolution
fibres and 132 MEDUSA mid-resolution fibres. For the UVES spectra, discussed in details in Mu13, 
we employed the 580 Red Arm set-up, with spectral resolution $R\sim$40000 and wavelength coverage $\sim$4800--6800 \AA .~
The GIRAFFE targets have been observed with the HR21 setup, with a resolving power of $\sim$16000 
and a spectral coverage between $\sim$8480--9000 \AA\ . This grating was chosen because it includes 
the prominent Ca~II triplet lines, which are ideal to measure radial velocities (RV) also in 
spectra of faint stars and to derive an estimate of their metallicity.

Two configurations of target stars have been used.
A total of 4 exposures of 46 min each for each configuration has been secured 
in Service Mode during the period between April and July 2012. A small overlap between the two configurations 
(12 stars) has been secured in order to cross-check the stability of the RV when measured 
with different fibres.

\begin{figure}
\includegraphics[width=84mm]{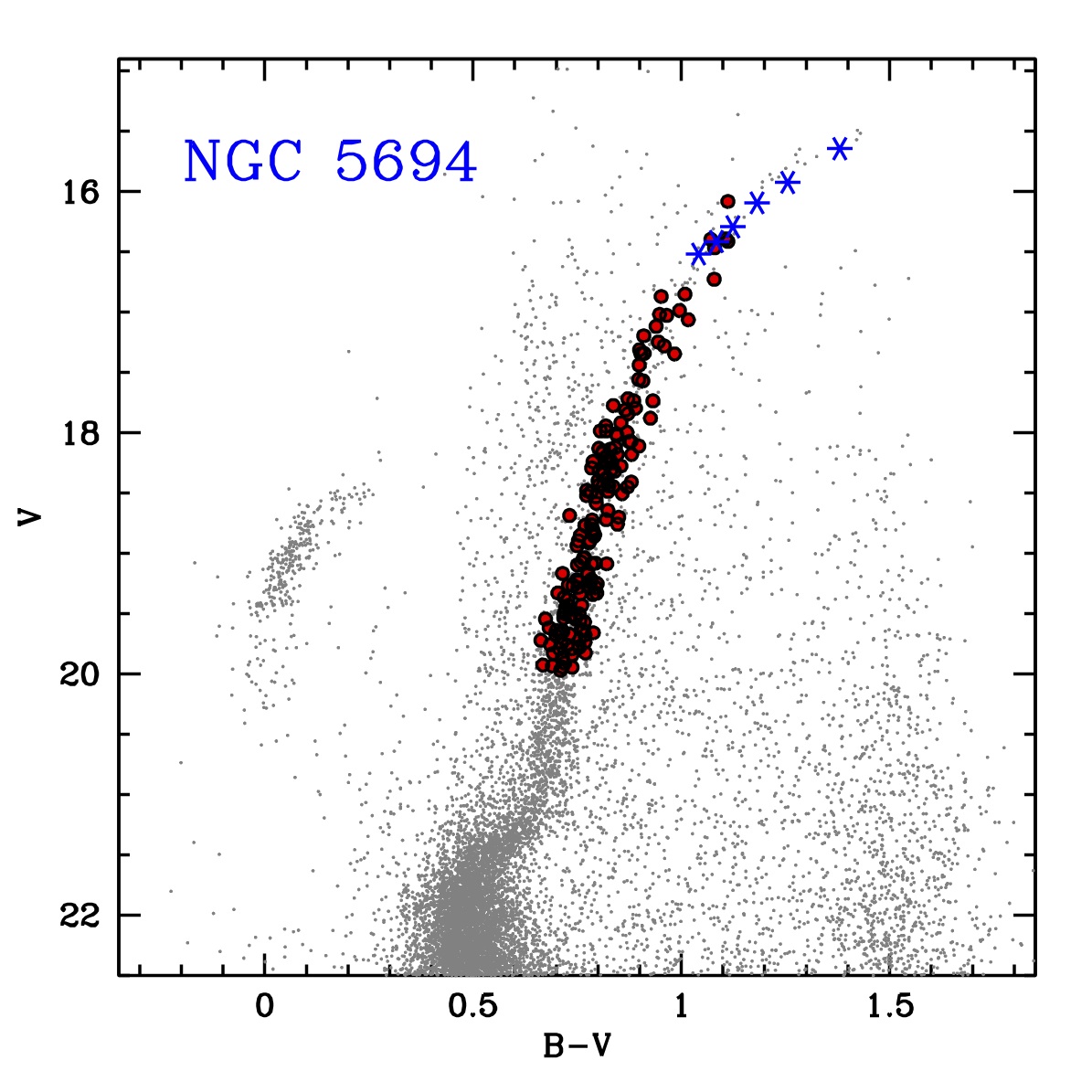}
\caption{Color Magnitude Diagram of NGC5694 in the V, B-V plane from VIMOS@VLT and WFPC2@HST 
(grey small points). Red circles and blue asterisks are the GIRAFFE and UVES targets, respectively.}
\label{cmd}
\end{figure}

\begin{figure}
\includegraphics[width=84mm]{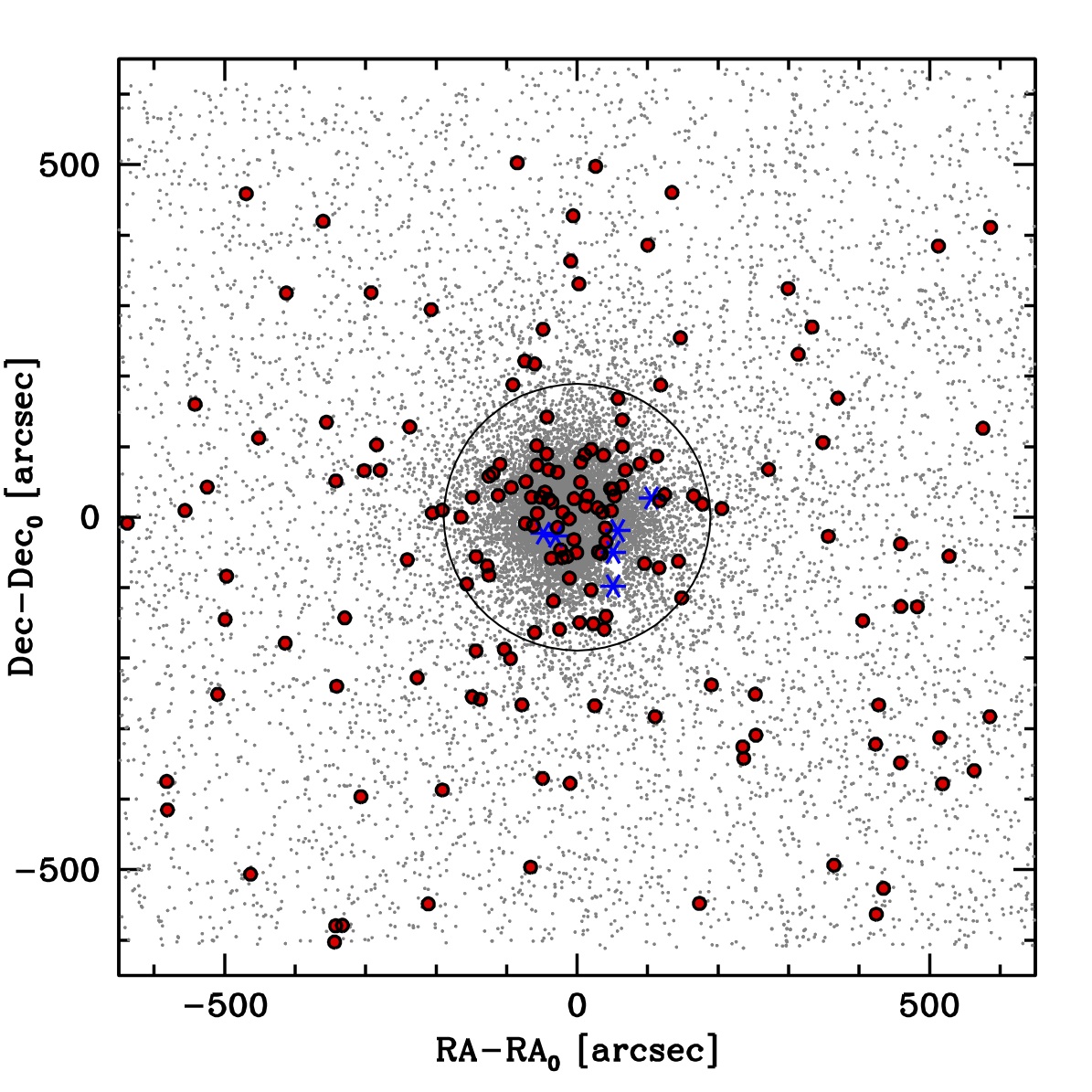}
\caption{Spatial distribution of the FLAMES targets (same symbols of Fig.~\ref{cmd}) 
with respect to the cluster center by \citet{noyola06}. The black circle
indicate the tidal radius of the K66 model providing the best-fit to the SB profile of NGC~5694 (from C11).}
\label{map}
\end{figure}

The target selection has been performed with the B,V photometric catalog by C11 obtained 
by combining VIMOS@VLT and WFPC2@HST data. We selected stars along the RGB with V$<$20.
%in order to 
%reach also for the faintest stars a SNR$\sim$10-15, still adequate to measure the Ca~II lines. 
Stars with close (within 2 arcsec) companion stars of comparable or brighter magnitude have been discarded, 
to avoid spurious contaminations in the fibre.
About 15-20 fibres in each configuration have been dedicated to sample the sky background, because 
this spectral range is affected by prominent $O_2$ and OH sky emission lines. 
Fig.~\ref{cmd} shows the (V, B-V) colour-magnitude diagram of NGC~5694 with 
with GIRAFFE targets marked as red circles and the UVES ones as blue asterisks. 
The spatial distribution of the targets 
with respect to the cluster center is shown in Fig.~\ref{map}; a circle with radius equal to the cluster tidal radius of the K66 model providing the best-fit to the SB profile of NGC~5694, as derived in C11, is also plotted, for reference.

%GEISLER???

%%%%% PRERIDUZIONE
The data reduction has been performed using the last version of the ESO pipeline
\footnote{http://www.eso.org/sci/software/pipelines/}, including 
bias-subtraction, flat-fielding, wavelength calibration with a standard Th-Ar lamp and 
spectral extraction. 
The accuracy of the zero-point of the wavelength calibration has been checked by measuring the position 
of several sky emission lines and comparing them with their rest-frame position taken from 
the sky lines atlas by \citet{oster96}. 
For each star the average difference between the measured and reference line positions 
is always smaller than 0.02 \AA, corresponding to less than one half of a pixel. 
These shifts turns out to be compatible with 0 within the quoted uncertainties. 
Hence, no relevant wavelengths shift is found.

Each individual stellar spectrum has been subtracted from the sky by using a master sky spectrum 
obtained as a median of the different sky spectra observed in that exposure. Then the proper 
heliocentric correction has been applied. Finally, individual spectra of the same target have been 
combined together.
The typical SNR per pixel (measured at $\sim$8550 $\AA$) is of $\sim$150 for the brightest targets 
(V$\sim$16) and of $\sim$10 for the faintest stars (V$\sim$20).
Some targets have been discarded because of the poor quality of their spectra or 
due to some residuals of the sky lines that can affect the correct RV measurement. 
Finally, the following analysis is based on 165 stars with reliable RV and metallicity estimates.

\section{Radial velocities}

RVs have been measured with the standard cross-correlation technique of the observed spectrum 
against a template of known RV, as implemented in the IRAF\footnote{{\sc iraf} (Image Reduction and Analysis Facility) is
distributed by the National Optical Astronomy Observatory, which is operated by
the Association of Universities for Research in Astronomy, Inc., under
cooperative agreement with the National Science Foundation.} task FXCOR. 
As template, we adopted a synthetic spectrum calculated with the code SYNTHE, adopting the entire 
atomic and molecular line-list by Kurucz and Castelli\footnote{http://wwwuser.oat.ts.astro.it/castelli/odfnew.html} 
and a ATLAS9 model atmosphere calculated with the metallicity of the cluster, [Fe/H]$\sim$--2.0 dex and 
the typical atmospheric parameters of a giant star ($T_{eff}$=~4500 K, log(g)=~1.5). 
Uncertainties in the RV have been computed by FXCOR according to \citet{tonry79}, 
by taking into account the height and width of the cross-correlation function 
and the mean distance between its main peak and the nearest secondary peaks \citep[Eq.~24 by][]{tonry79}.

As an additional sanity check, the RVs of the 12 stars observed in both configurations 
have been measured individually. The two sets of RV agree very well each 
other, with a mean difference of $+0.5\pm 0.5$~ km/s, fully compatible with 
a null difference. 

\begin{figure}
\includegraphics[width=84mm]{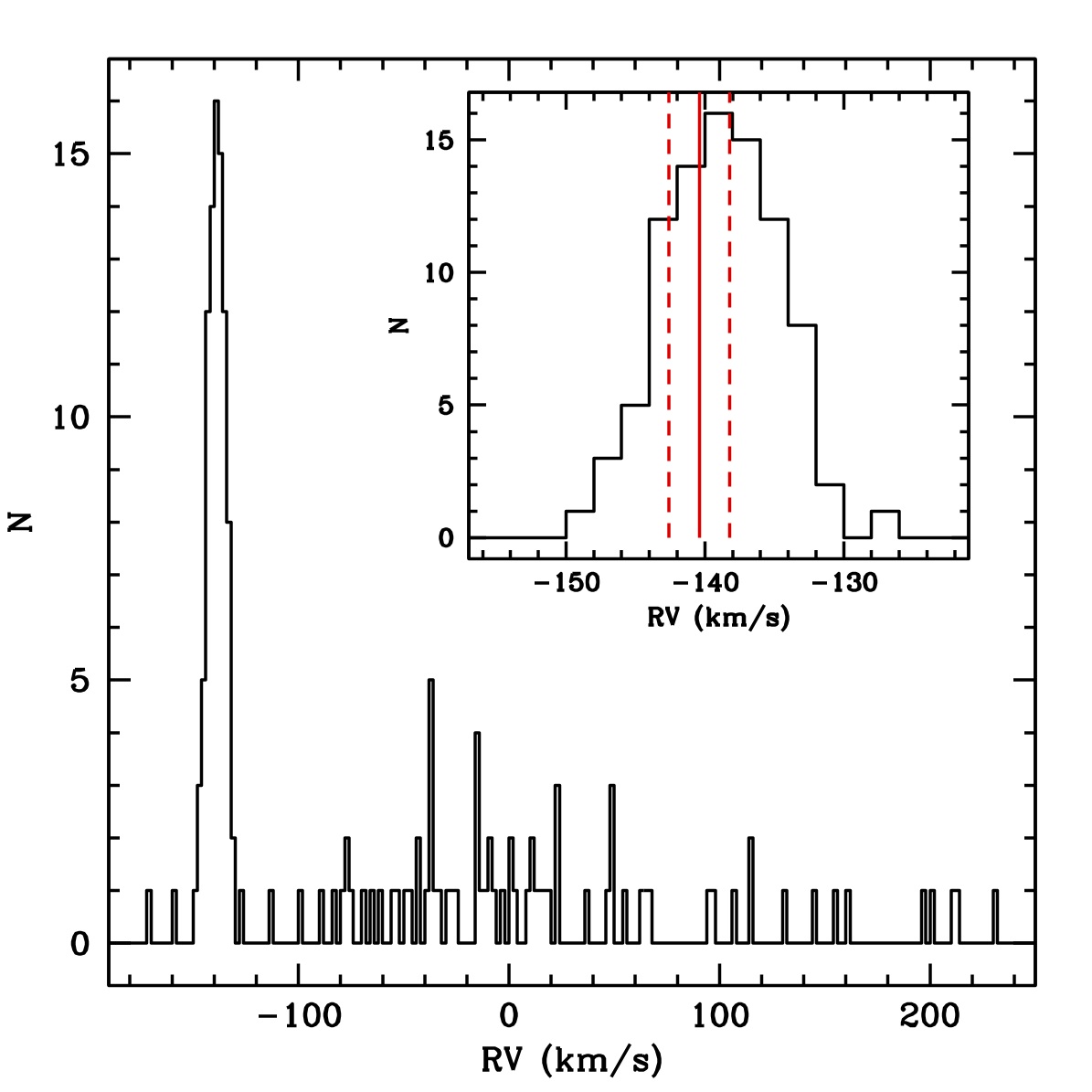}
\caption{RV distribution of the stars of NGC~5694 observed with GIRAFFE. The inset panel 
shows a zoomed view of the region around the main peak of the distribution. 
The red solid line is the average RV obtained from the six UVES targets Mu13, 
while the two dashed lines indicate $\pm1\sigma$ levels.}
\label{rvdist}
\end{figure}

Fig.~\ref{rvdist} shows the RV distribution of the entire sample of GIRAFFE targets. 
The distribution ranges from --171 km/s to +231 km/s, with a dominant peak around at $\sim$--140 km/s 
and corresponding to the cluster stars (the average RV derived from the 6 UVES targets is of 
-140.4$\pm$2.2 km/s). The inset panel shows a zoomed view of the region around the main peak of the 
RV distribution, with marked as a reference the average RV obtained from the UVES targets 
(red solid vertical line).

%%%%%%%%%%%%%%%%%%%%%%%%%%%%%%%%%%%%%%%%%%%%%%%%%%%%%%%%%%%%%%%%%%%%%%%%%%%%%%
%%%%%%%%%%%%%%%%%%%%%%%%%%%%%%%%%%%%%%%%%%%%%%%%%%%%%%%%%%%%%%%%%%%%%%%%%%%%%%

\section{Metallicity}

%%%%
\subsection{Abundances from the Ca~II triplet lines}
Abundances for all the target stars have been obtained by using the strength of the 
Ca~II triplet lines as a proxy of the metallicity. 
The lines of the Ca~II triplet lines have been fitted
with a Voigt profile, in order to reproduce the prominent pressure-broadened line wings, 
and then their equivalent widths (EWs) obtained by direct integration of the best-fit profile.
The metallicities have been obtained by adopting the calibration by \citet{carrera07} and 
assuming $V_{HB}$=~18.5 mag \citep{harris10}. 

Uncertainties in the measured EWs of the Ca~II triplet lines have been estimated by 
employing Monte Carlo simulations. A synthetic spectrum has been re-sampled 
at the pixel-size of the GIRAFFE spectra (0.05 \AA\ /pixel) and then Poissonian noise 
corresponding to four values of SNR (namely 10,50,100,150) has been injected in order to simulate 
the noise conditions of the observed spectra. For each value of SNR, 1000 synthetic spectra 
have been generated following this approach, the EWs of the Ca~II triplet lines have been measured 
and then the metallicity estimated as done for the observed spectra. 
For each SNR, the uncertainty has been computed  as 1$\sigma$ of the derived 
[Fe/H] distribution; the abundance uncertainty is of 0.13 dex for SNR=~10 and 0.01 dex for SNR=~150.
A relation that provides $\sigma_{[Fe/H]}$ as a function of SNR has been derived and
used to estimate the uncertainty in [Fe/H] of all the targets interpolating at their value of the SNR.
Additional sources of uncertainty are the error in V-$V_{HB}$
(a variation of $\pm$0.05 mag translates in a variation in [Fe/H] of $\mp$0.01 dex) and 
the uncertainty in the $EW_{CaT}$--[Fe/H] linear fit \citep[$\sigma$=~0.08 dex, as quoted by][]{carrera07}.
The distribution of the [Fe/H] abundance ratios as derived from Ca~II triplet lines is shown in 
Fig.~\ref{iron} as an empty histogram.

%%%%

\subsection{Abundances from the Fe~I lines}
For 19 targets, the quality of the spectra (SNR$>$40) allows to determine the iron abundance 
directly from the measure of Fe~I lines. We identify a ten of Fe~I lines, unblended 
at the HR21 setup resolution and at the atmospheric parameters and metallicity of the 
targets (see Mu13, for details). 
EWs measurements have been performed by using the code DAOSPEC \citep{stetson}, 
iteratively launched 
by means of the package 4DAO\footnote{http://www.cosmic-lab.eu/4dao/4dao.php}\citep{m13_4dao} 
that allows an analysis cascade of a large sample of stellar spectra and a visual inspection 
of the Gaussian fit obtained for all the investigated lines.

The iron abundance has been derived with the package 
GALA\footnote{http://www.cosmic-lab.eu/gala/gala.php} \citep{m13g}, by matching the measured 
and the theoretical EWs.
Atmospheric parameters have been derived by using 
the B,V photometry by C11. 

$T_{eff}$ have been computed by means of the $(B-V)_0$--$T_{eff}$ 
transformation by \citet{alonso99} based on the Infrared Flux Method; the de-reddened color 
$(B-V)_0$ is obtained adopting a color excess E(B-V)=~0.099 mag (C11) and the 
extinction law by \citet{mccall}. 
Surface gravities have been computed with the Stefan-Boltzmann relation, assuming the 
photometric $T_{eff}$, the distance modulus of 17.75$\pm$0.10 mag (C11) and an evolutive mass 
of 0.75 $M_{\odot}$, according to an isochrone from the BaSTI dataset \citep{pietrinferni04} 
with age of 12 Gyr, Z=~0.0003 and a solar-scaled chemical mixtures.
The bolometric corrections are calculated according to Eq.~(17) of \citet{alonso99}.
Micro-turbulent velocities cannot be estimated from these spectra because of the small number of available 
lines and we adopted for all the targets the average value obtained by the UVES spectra, $v_{turb}$=~1.8 km/s.

\begin{figure}
\includegraphics[width=84mm]{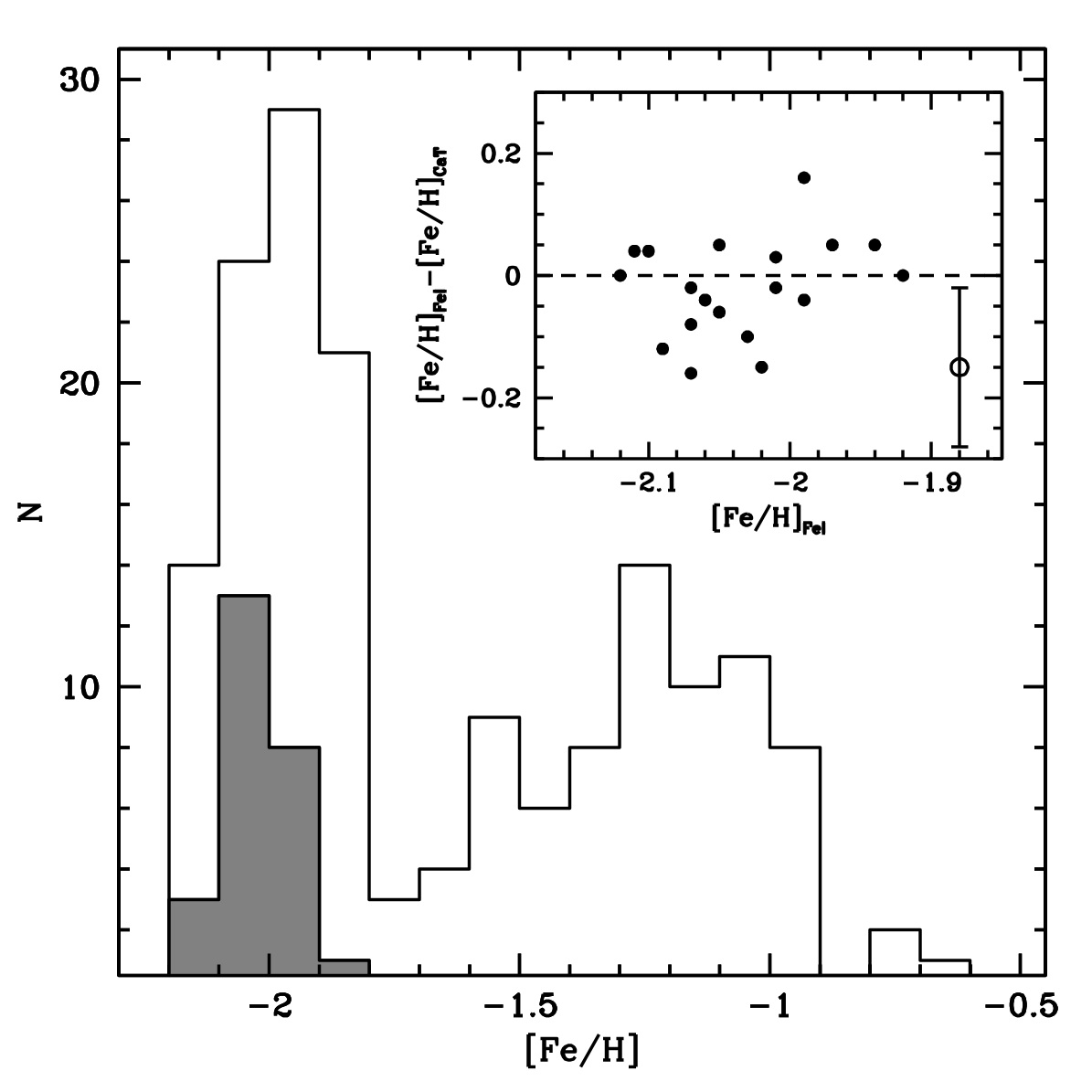}
\caption{[Fe/H] distribution of the targets as obtained from the Ca~II triplet lines 
(empty histogram) and from the direct measurement of Fe~I lines (UVES+GIRAFFE, grey-shaded histogram). 
The inset panel show the difference between the iron content from Fe~I lines and from Ca~II triplet lines 
as a function of $[Fe/H]_{FeI}$ for 19 target stars observed with FLAMES. The typical uncertainty in the difference is shown as the error-bar of the empty circle in the lower-right corner of the inset.}
\label{iron}
\end{figure}

The average iron abundance of these 19 stars is [Fe/H]=--2.04$\pm$0.02 dex, in reasonable agreement with 
the value derived by the UVES spectra, [Fe/H]=--1.98$\pm$0.03 dex (Mu13).  
Also, we highlight the good agreement between the abundances derived from Ca~II triplet and from 
Fe~I lines: the average difference of the iron abundances for the 19 stars in common is 
$[Fe/H]_{CaT}-[Fe/H]_{Fe}=+0.02\pm 0.01$ dex ($\sigma$=~0.08 dex). 
Fig.~\ref{iron} shows the distribution of the 19 stars as a grey histogram, while the inset panel shows 
the difference between [Fe/H] from Fe~I lines and from Ca~II lines.

We consider together the 25 stars for which a direct Fe abundance has been derived, 
19 GIRAFFE and 6 UVES targets. The mean abundance, together with the intrinsic spread $\sigma_{int}$ 
and their uncertainties, have been calculated with the maximum likelihood algorithm 
described in \citet{m12}. 
We obtain an average value of [Fe/H]=--2.01$\pm$0.02 dex ($\sigma_{int}$=~0.0$\pm$0.03 dex) that 
we recommended as the final value for the Fe abundance. 
In comparison, the same algorithm applied on the [Fe/H] derived from Ca~II triplet lines of the 83 member stars 
provides [Fe/H]=--1.99$\pm$0.01 dex ($\sigma_{int}$=~0.0$\pm$0.02 dex).

%%%%%%%%%%%%%%%%%%%%%%%%%%%%%%%%%%%%%%%%%%%%%%%%%%%%%%%%%%%%%%%%%%%%%%%%%%%%%%
%%%%%%%%%%%%%%%%%%%%%%%%%%%%%%%%%%%%%%%%%%%%%%%%%%%%%%%%%%%%%%%%%%%%%%%%%%%%%%

%%%%%%%%%%%%%%%%%%%%%%%%%%%%%%%%%%%%%%%%%%%%%%%%%%%%%%%%%%%%%%%%%%%%%%%%%%%%%%
%%%%%%%%%%%%%%%%%%%%%%%%%%%%%%%%%%%%%%%%%%%%%%%%%%%%%%%%%%%%%%%%%%%%%%%%%%%%%%
\begin{table*}
\begin{minipage}{145mm}
\caption{Main parameters for target stars observed with FLAMES: identification number, right ascension and declination, SNR per pixel,  B and V 
magnitudes (C11), radial velocity, [Fe/H] from direct Fe~I lines measurement and from Ca~II triplet lines. 
Uncertainties in $[Fe/H]_{CaT}$ include both internal errors and the uncertainty in the calibration by \citet{carrera07}.
The entire table is available in the electronic version of the journal.}
\begin{tabular}{lcccccccc}
\hline
   ID &         RA    &  Dec          &  SNR   & B & V &  RV        & $[Fe/H]_{Fe}$ & $[Fe/H]_{CaT}$ \\
      &      (J2000)  &  (J2000)      & (@8550\AA) &   &   &  (km/s)    & (dex)	    &	(dex) \\
\hline
  65 &   219.8460466 &  -26.5338128  & 150  &  17.194  &    16.082   &    -33.15$\pm$0.52  & ---	      & --1.07$\pm$0.08  \\
  88 &   219.8965614 &  -26.5306057  & 140  &  17.498  &    16.391   &   -137.78$\pm$0.21  & --1.92$\pm$0.11  & --1.92$\pm$0.08  \\
  89 &   220.0596065 &  -26.6086914  &  80  &  17.467  &    16.396   &     -8.53$\pm$0.66  & ---	      & --1.39$\pm$0.08  \\
  94 &   219.8884807 &  -26.5485943  &  99  &  17.525  &    16.413   &   -138.75$\pm$0.21  & --1.97$\pm$0.11  & --2.02$\pm$0.08  \\
  99 &   219.9004115 &  -26.4470219  &  68  &  17.546  &    16.466   &    -10.71$\pm$0.58  & ---	      & --1.29$\pm$0.08  \\
\hline
\end{tabular}
\end{minipage}
\end{table*}
%%%%%%%%%%%%%%%%%%%%%%%%%%%%%%%%%%%%%%%%%%%%%%%%%%%%%%%%%%%%%%%%%%%%%%%%%%%%%%
%%%%%%%%%%%%%%%%%%%%%%%%%%%%%%%%%%%%%%%%%%%%%%%%%%%%%%%%%%%%%%%%%%%%%%%%%%%%%%

\section{Kinematic analysis}

\subsection{Membership}

\begin{figure}
\includegraphics[width=84mm]{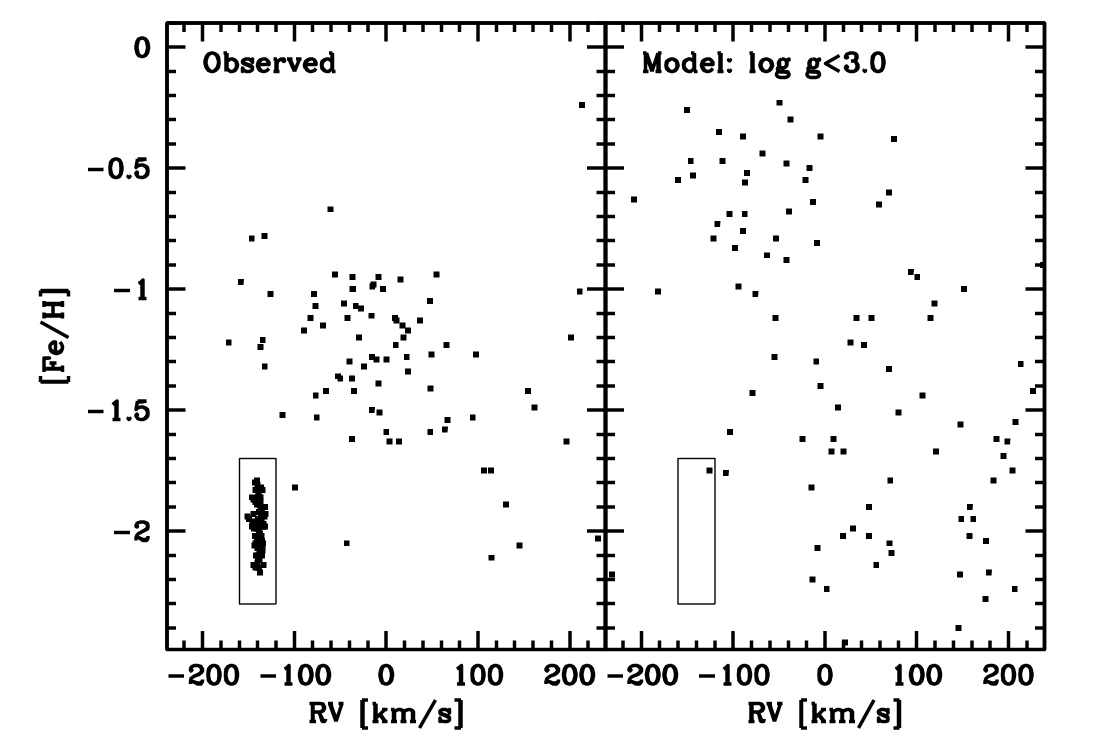}
\caption{Comparison of our sample (left panel) with the predictions of the Besancon Galactic model (right panel) 
in the RV vs metallicity plane. The model stars are those predicted for a field of $1\degr \times 1 \degr$ in the 
direction of the cluster. We included in this plot only stars in a color - magnitude window enclosing our targets 
($0.6<B-V<1.4$ and $15.5<V<20.0$) and in the same range of surface gravity as cluster RGB stars (log g$<3.0$). 
The thin rectangle is the box that we adopted to select our bona-fide cluster members.}
\label{fe_rv}
\end{figure}

Fig.~\ref{fe_rv} compares the distribution of the RV as a function of [Fe/H] 
%(considering the abundances from Ca~II triplet lines for the GIRAFFE targets 
%and those from Fe~I lines for the UVES targets) 
of our targets with that predicted by the Besancon Galactic Model \citep{robin03}
for a $1\degr \times 1 \degr$ field in the direction of NGC~5694. In the right panel
of Fig.~\ref{fe_rv} we plot only model stars with surface gravity in the range covered
by cluster RGB stars (log(g)$<3.0$) and lying within a window in the CMD that encloses
our targets ($0.6<B-V<1.4$ and $15.5<V<20.0$). Stars belonging to the cluster are very clearly identified
in this plane, forming a tight and isolated concentration around $V_r\sim -140$~km/s and [Fe/H]$\sim -2.0$.
For this reason, we adopted the rectangular box plotted in both panels of the figure to select 
our sample of candidate cluster members. Target stars are selected as cluster members if they satisfy 
the conditions -160.0~km/s$<V_r<$-120.0~km/s and $-2.3<$[Fe/H]$<-1.7$. It is reassuring to note that only 
one over 101 model stars falls within the selection box; given the difference in the sampled area, this corresponds 
to a probability $<0.2$\% to have a Galactic giant contaminating our sample. If also dwarf stars are considered (log g$\ge 3.0$), 
only 1 model star over 5074 falls in the box. We can conclude that our selection criteria are robust and that all the selected 
stars can be considered as bona-fide cluster members\footnote{It must be noted that [Fe/H] estimates from CaT for 
non-member stars are by definition not correct, since the $V-V_{HB}$ parameter is ill-defined for stars lying at any distance from the cluster.
A variation of $\pm$0.5 mag in the adopted distance modulus leads to a variation of $\mp$0.13 dex in the derived [Fe/H]. 
This behaviour is consistent with the different slopes in the [Fe/H]--RV plane found for the observed stars and the 
Besancon Galactic Model.}.

According to these criteria, we selected 83 out 165 GIRAFFE targets. 
In the following we use the selected sample of 89 bona-fide members (83 GIRAFFE plus 6 UVES targets), located
between $0.2\arcmin$ and $6.5\arcmin$ from the cluster center,  
to study the kinematics of the cluster. The mean RV value, computed with a 
maximum likelihood algorithm \citep[ML hereafter; see][]{walker06}, is --139.2$\pm$0.4 km/s, 
in good agreement with previous estimates \citep{geisler95,dubath97,lee}.

%%%%%%%%%%%%%%%%%%%%%%%%%%%%%%%%%%%%%%%%%%%%%%%%%%%%%%%%%%%%%%%%%%%%%%%%%%%%%%
%%%%%%%%%%%%%%%%%%%%%%%%%%%%%%%%%%%%%%%%%%%%%%%%%%%%%%%%%%%%%%%%%%%%%%%%%%%%%%

\subsection{Cluster rotation}
We searched for rotation adopting the method and the notation described 
in \citet{bellazzini12}. 
Basically, the sample is divided into two groups by a line 
passing from the cluster center, and the difference between 
the average RV of the two sub-samples, on each side of the dividing line, 
is computed. This step is repeated by varying 
the value of the position angle (PA) of the boundary line in steps of 10$^{\circ}$.

\begin{figure}
\includegraphics[width=84mm]{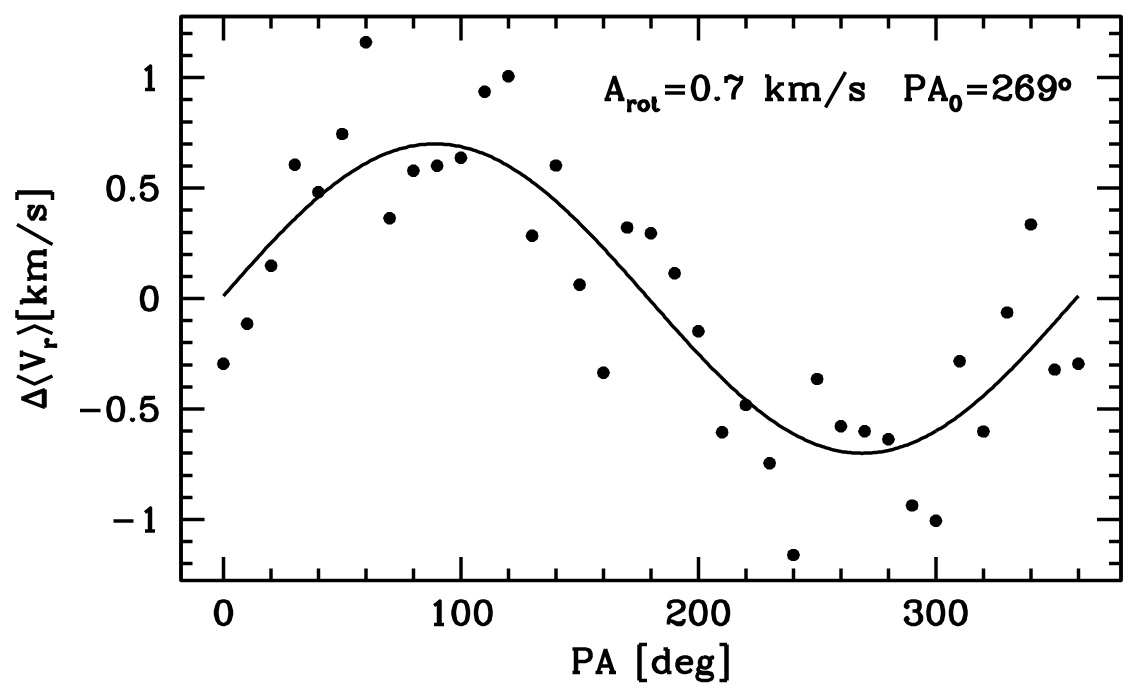}
\caption{Rotation as derived from the FLAMES sample of NGC~5694. 
The difference between the average RV on each side of the cluster 
with respect to a line passing through the cluster center with a 
given position angle (PA) is shown as a function of the PA itself. 
The continuos line is the sine function that best fits the observed pattern,
$A_{rot}$ and $PA_0$ the best-fit amplitude and position angle \citep[see][]{bellazzini12}.}
\label{rot}
\end{figure}

The difference between the mean RV of the two sub-samples as a function of PA 
is shown in Fig~\ref{rot}, together with 
the sine function that best fits the observed pattern. 
The best-fit sine function has a position angle of the rotation axis of 269$^{\circ}$ 
and an amplitude of the rotation curve of 0.7~km/s\footnote{$A_{rot}$ is in fact the maximum difference between the mean velocity in the two considered halves of the cluster. This is two times the mean rotation amplitude in the considered radial range. \citet{bellazzini12} argue that in many cases $A_{rot}$ is a reasonable proxy for the actual maximum amplitude.} According to \citet{bellazzini12} 
this very weak amplitude of the mean rotation is typical of clusters as metal-poor as (and with an Horizontal Branch - HB - morphology as blue as) NGC~5694.
Since the rotation amplitude is significantly lower than the velocity dispersion over the
whole radial range covered by our data (see below) it can be neglected 
in the following analysis.

%%%%%%%%%%%%%%%%%%%%%%%%%%%%%%%%%%%%%%%%%%%%%%%%%%%%%%%%%%%%%%%%%%%%%%%%%%%%%%
%%%%%%%%%%%%%%%%%%%%%%%%%%%%%%%%%%%%%%%%%%%%%%%%%%%%%%%%%%%%%%%%%%%%%%%%%%%%%%

\subsection{Velocity dispersion}

The projected velocity dispersion profile has been derived following 
the same procedure described in \citet{b08}. 
The cluster area has been divided in 4 concentric annuli, in order 
to have in each radial bin a similar number of stars ($\sim$20-24).
In each radial bin the velocity dispersion $\sigma_{RV}$ and the
associated errors ($\epsilon_{\sigma}$) have been 
computed with the ML method \citep{walker06}, keeping the systemic velocity fixed. 
An iterative 3$\sigma$ clipping algorithm applied in each radial bin
did not lead to the rejection of any additional star. The derived
profile is reported in Table~2.

In the following we will compare the observed velocity dispersion profile
of NGC~5694 with different kind of theoretical models. A detailed assessment
of the best model, as performed, e.g., in \citet{iba11}, is beyond the scope of the present
analysis, and is also prevented by our lack of the full control of uncertainties in the 
composite SB profile by C11, that is required for that kind of analysis.
On the other hand, our main purpose is to explore models that can provide
a reasonable representation of both the SB and the velocity dispersion profile
to get insight on the physical characteristics of this anomalous stellar system,
within the boundaries of Newtonian dynamics.

\subsubsection{Comparison with single-mass isotropic models}

\begin{figure}
\includegraphics[width=84mm]{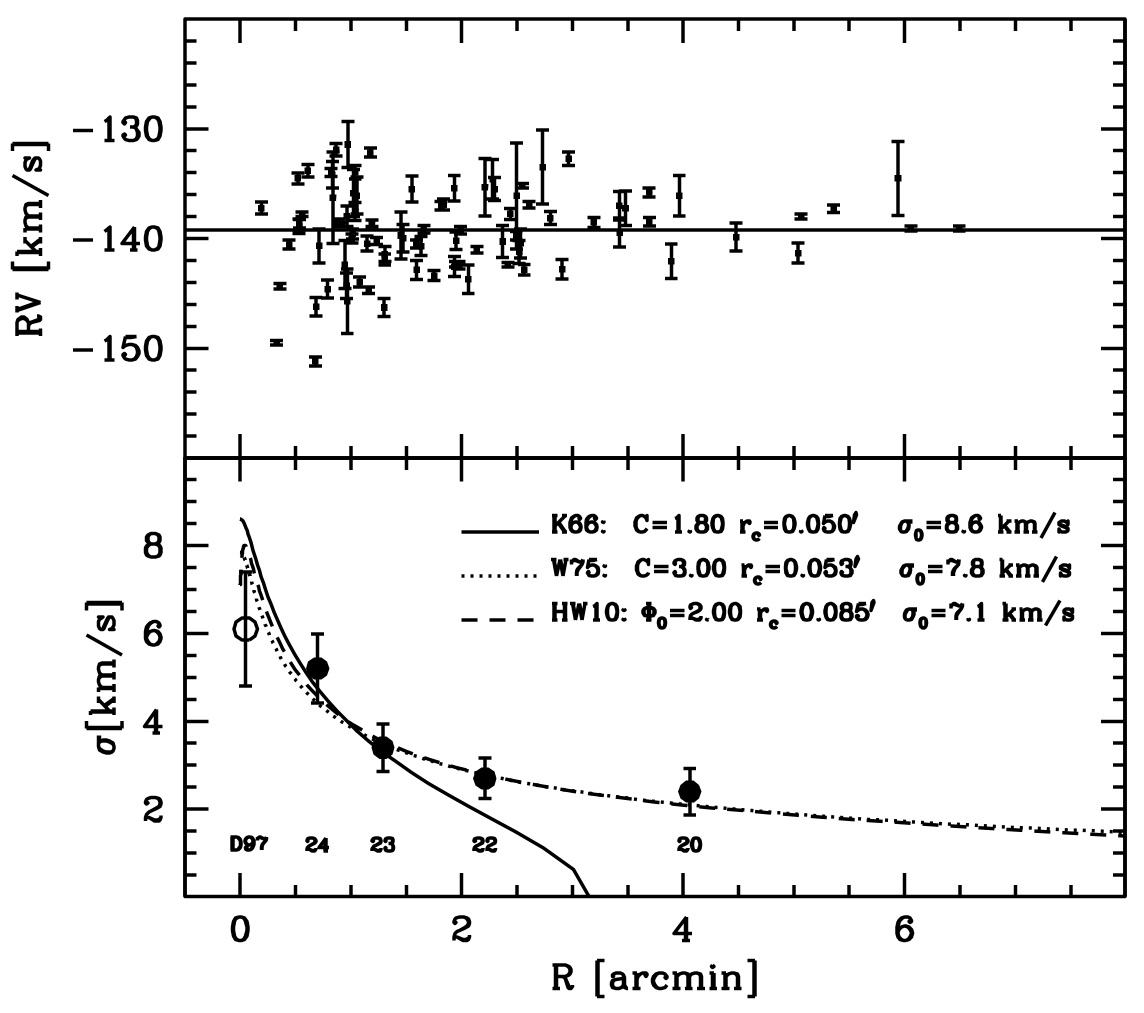}
\caption{Upper panel:
RV distribution of the individual member stars as a function of the distance 
from the cluster center. Solid horizontal line indicates the systemic RV 
of the cluster.
Lower panel:
velocity dispersion as a function of the distance from the cluster center 
from Table 2 (black circles). The open circle is 
the central velocity dispersion estimate from integrated spectroscopy by \citet{dubath97}.
The number of stars per bin is also labelled. 
Theoretical models are over-imposed as comparison, namely by
\citet[][solid curve]{king66}, \citet[][dotted curve]{wilson} and 
\citet[][dashed curve]{hw10}. The parameters of the models adopted for the fit
are also reported.}
\label{disp}
\end{figure}

The upper panel of Fig.~\ref{disp} shows the RV distribution as a function of 
the distance from the cluster center \citep[assuming the coordinate of the center from][]{noyola06} 
for the individual member stars. The lower panel of Fig.~\ref{disp} shows the derived 
velocity dispersion profile, where the black dots are the values of $\sigma_{RV}$ 
derived in each radial bin from our data.
Our profile is complemented by the central value provided 
by \citet[][$\sigma_{0}$=~6.1$\pm$1.3~km/s]{dubath97}, obtained 
from integrated spectroscopy (empty point). 
The velocity dispersion gently declines from the center
to $\sim 2\arcmin \simeq 7.1 r_h$\footnote{Where the observed half-light radius is $r_h=0.28\arcmin$, from C11.}, and then flattens out to $\sigma\simeq 2.5$~km/s in the two outermost bins.
Note that the outermost point of our velocity dispersion profile lies at $\simeq 4\arcmin$, 
corresponding to more than 14$r_h$ from the center, and still is far away from the limits of 
the cluster, since in C11 we were able to trace the SB profile out to $R=8.5\arcmin\simeq 30 r_h$.

%%%%%%%%%%%%%%%%%%%%%%%%%%%%%%%%%%%%%%%%%%%%%%%%%%%%%%%%%%%%%%%%%%%%%%%%% TABELLA VEL DISP PROF
\begin{table}
\label{tab_sig}
 \centering
 \begin{minipage}{70mm}
  \caption{Velocity dispersion profile.}
  \begin{tabular}{@{}cccccc@{}}
  \hline
    $R_{in}$  & $R_{out}$   &  $\langle R\rangle$ & $\sigma_{RV}$ & $\epsilon_{\sigma}$ & $N_{star}$\\
 $[$arcmin$]$ &$[$arcmin$]$ &     $[$arcmin$]$    &  [km/s]       &    [km/s]           &            \\
  \hline
0.0 & 1.0 &  0.70 &   5.2 &	   0.78 &  24 \\  
1.0 & 1.8 &  1.29 &   3.4 &	   0.54 &  23 \\  
1.8 & 2.6 &  2.21 &   2.7 &	   0.46 &  22 \\  
2.6 & 7.8 &  4.06 &   2.4 &	   0.53 &  20 \\  
\hline
\end{tabular}
\end{minipage}
\end{table}
%%%%%%%%%%%%%%%%%%%%%%%%%%%%%%%%%%%%%%%%%%%%%%%%%%%%%%%%%%%%%%%%%%%%%%%%% FINE TABELLA VEL DISP PROF

%%%%%%%%%%%%%%%%%%%%%%%%%%%%%%%%%%%%%%%%%%%%%%%%%%%%%%%%%%%%%%%%%%%%%%%%%%%%%%%
%%%%%%%%%%%%%%%%%%%%%%%%%%%%%%%%%%%%%%%%%%%%%%%%%%%%%%%%%%%%%%%%%%%%%%%%%%%%%%%

%%%%%%%%%%%%%%%%%%%%%%%%%%%%%%%%%%%%%%%%%%%%%%%%%%%%%%%%%%%%%%%%%%%%%%%%%%%%%%%
\begin{figure}
\includegraphics[width=84mm]{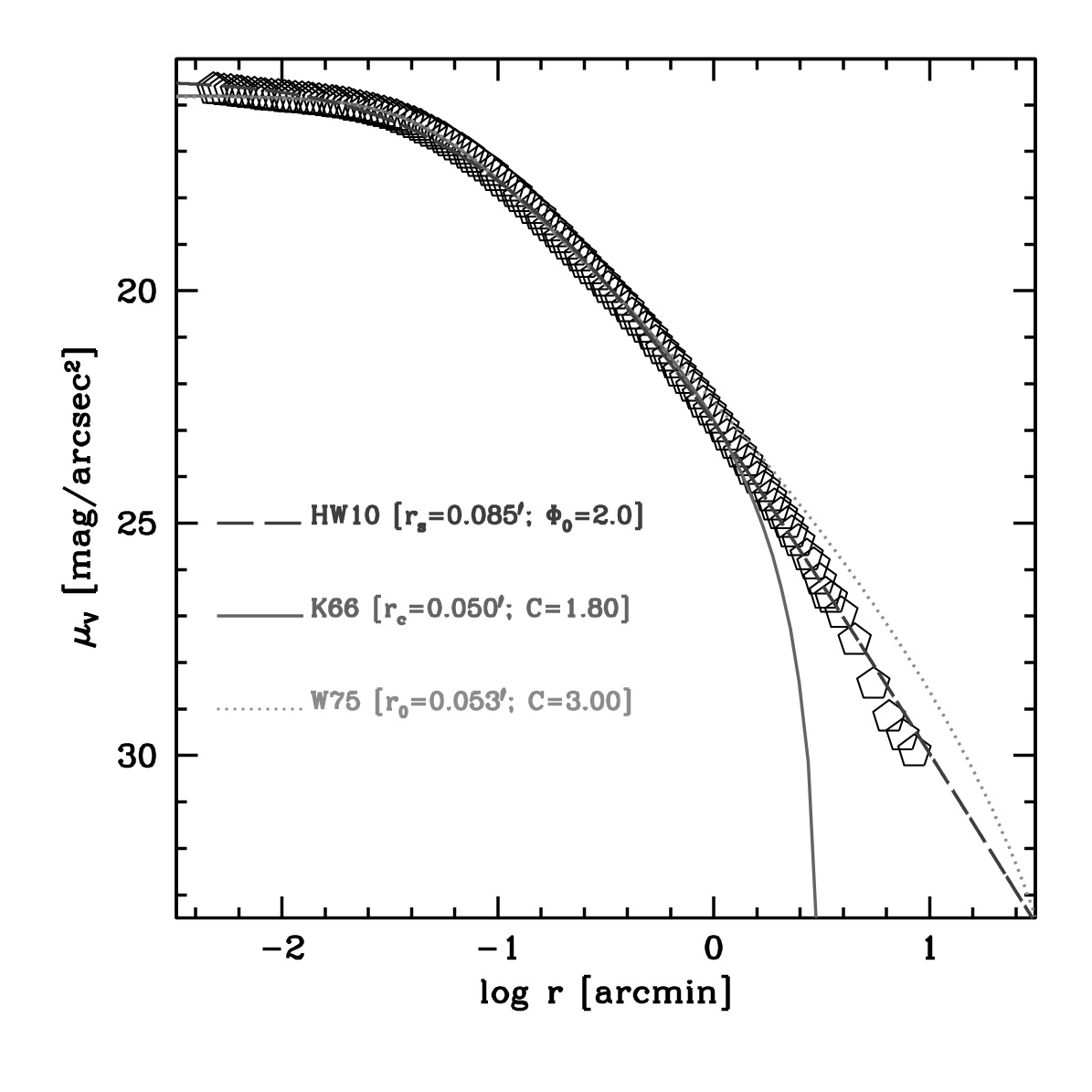}
\caption{Surface brightness profile of NGC~5694 from C11. In addition to the single-mass 
isotropic \citet{king66} 
and \citet{wilson} models that best-fits the profile already shown in C11, we superimpose also
a DARKexp model \citep{hw10} providing an acceptable fit over the whole extension of the observed profile.}
\label{SBprof}
\end{figure}
%%%%%%%%%%%%%%%%%%%%%%%%%%%%%%%%%%%%%%%%%%%%%%%%%%%%%%%%%%%%%%%%%%%%%%%%%%%%%%%

In the lower panel of Fig.~\ref{disp} we superimpose to the data the
predictions of three models
that fits the observed SB profile of the cluster (at least in the innermost regions, see C11 and Fig.~\ref{SBprof}). The models have been normalised to best fit the
observed velocity
dispersion profile. The first two models are the single-mass isotropic
\citet[][K66 hereafter]{king66} and \citet[][W75]{wilson}
models that were proposed in C11. In addition to these we adopt also a
{\em DARKexp} model \citep[][HW10]{hw10} that is shown,
in Fig.~\ref{SBprof}, to provide a reasonable fit of the SB profile over
the whole extension of the cluster, in particular
in the outermost region of the profile where K66, W75 and \citet{elson}
models fail (see C11)\footnote{It is interesting to note that also
models with a larger number of free parameters than those considered
here, like the Nuker or the core-Sersic models \citep[see][]{graham}, appear
unable to provide a satisfactory fit to the entire SB profile of NGC~5694.}. 
DARKexp models are theoretically
derived maximum entropy equilibrium states of self-gravitating 
collision-less systems (HW10). \citet{WBH} have
shown that in many cases (including NGC~5694)
they provide a better fit to the observed profiles of GCs, with respect to
K66 models.
The main difference between K66/W75 models and DARKexp models
is in the assumed Distribution
Function (DF; $f(E)\equiv \frac{dN}{d^3r~d^3v}$). K66 and W75 models adopt a
DF from the family of lowered Maxwellian distributions, which reproduces
the trend toward an isothermal condition driven by collisions. DarkEXP models,
instead, adopt a DF such that the corresponding energy distribution
(N(E)=$f(E)\int r^2 v dv$)
is a lowered Maxwellian distribution, with a proper treatment of the
low-occupation-number regime \citep[see][for details and
discussion]{WBH}.

The comparisons displayed in the lower panel of Fig.~\ref{disp} clearly
show the inadequacy of K66 models to describe
the kinematics of NGC~5694. This is due to the tidal truncation that is
built-in in K66 models: the SB profile is
unable to fit the extended $\sim R^{-3}$ outer profile of the cluster and,
consequently, it lacks sufficient mass
in the outer regions to sustain a (nearly) flat dispersion profile beyond
$R\sim 2\arcmin$. The lack of a tidal truncation may be related to 
the fact that the cluster is under-filling its Roche lobe by a significant amount
(see C11 and Sect.~\ref{sum}, for further discussion).
On the other hand both
the W75 and DARKexp models, that have much more extended SB profiles,
provide a fair representation of the cluster
kinematics.

The central dispersion estimate by \citet{dubath97} appears slightly low
with respect to our innermost point and
the extrapolation of all best-fitting models. It would be valuable to
have an independent estimate of the dispersion in the innermost regions
based on the velocities of individual stars, to obtain a robust validation
of the estimate by \citet{dubath97} from integrated spectroscopy. Moreover,
it has to be recalled that in this comparison we considered
only isotropic single-mass and non-collisional models. In the
innermost regions of the cluster collisional processes can be
important and mass-segregation, as well as anisotropy, is expected to contribute in shaping the
overall line-of-sight velocity dispersion profile.
%In particular multi-mass model tend to have lower central peaks in the velocity dispersion profile with respect to single-mass ones (see below).

\subsubsection{Comparison with multi-mass and anisotropic models}

\begin{figure}
\includegraphics[width=84mm]{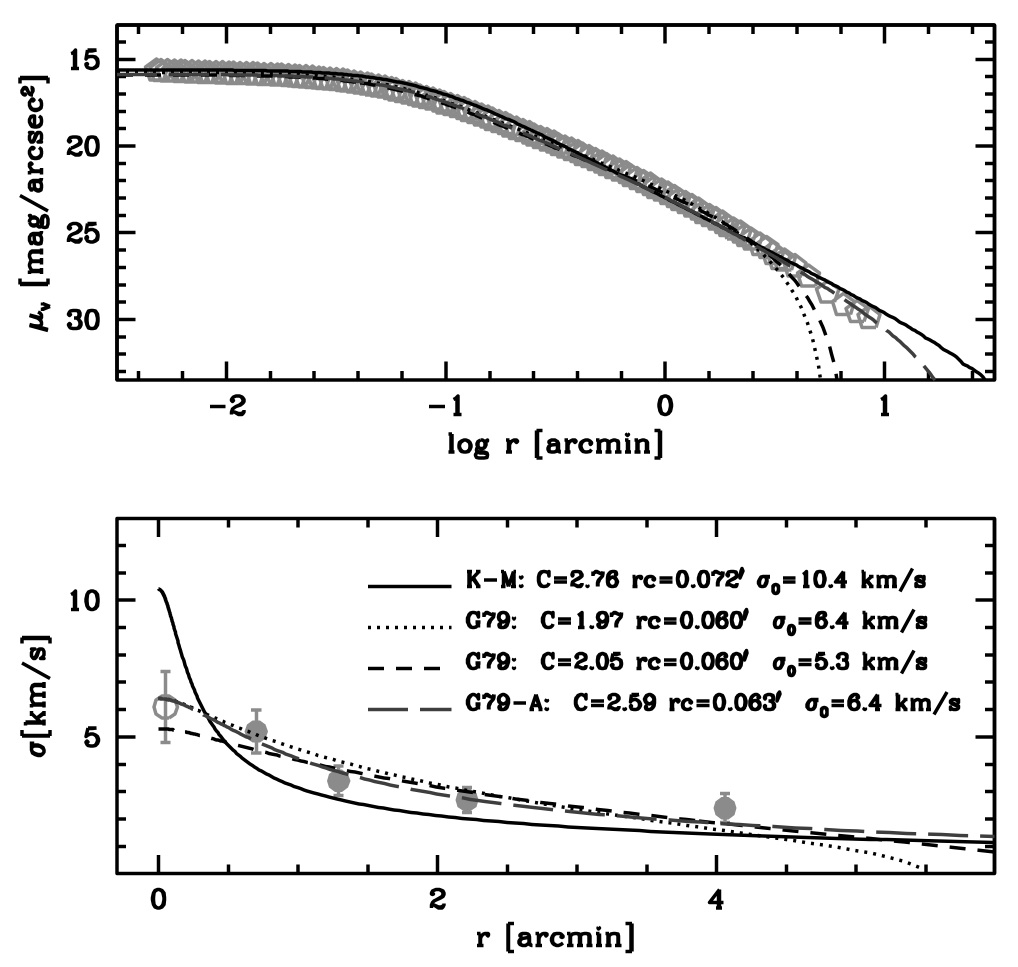}
\caption{The surface brightness profile (upper panel) and the velocity
dispersion profile (lower panel) of NGC~5694 are compared with a
single-mass King-Michie model with radial anisotropy ($2T_r/T_t=1.525$;
continuous line), and with the predictions for the distribution of star in
the largest mass bin (approximately corresponding to cluster giants) of
two multi-mass King models computed following G79.
The dotted lines correspond to a model with a single power law 
MF with index $x=-1$; the
short-dashed lines correspond to a model with \citet{kroupa} MF.
Finally, the long-dashed dark grey lines correspond to a multi-mass
model with \citet{kroupa} MF and radial anisotropy.
All the models have been normalised to best fit the observed velocity
dispersion profile.}
\label{aniso}
\end{figure}

To explore the possible role of orbital anisotropy and mass segregation in
the dynamics of NGC~5694 we compare, in Fig.~\ref{aniso}, the observed SB
and $\sigma$ profiles of the clusters with the predictions of a model
including radial anisotropy, two multi-mass models and one multi-mass anisotropic model.

The continuos lines in both panels of Fig.~\ref{aniso} correspond to a
single-mass King-Michie \citep[K-M][]{michie} model with  the maximum
degree of radial anisotropy that still ensures stability
\citep[see][for a thorough discussion of this family of models]{iba11}.
 In these models, orbits are isotropic in the center and becomes
radially biased at a characteristic radius $r_{a}$.
A rough criterion for stability is represented by the so-called
Fridman-Polyachenko-Shukhman parameter $\xi=2T_{r}/T_{t}$ (Friedman \&
Polyachenko 1984),
where $T_{r}$ and $T_{t}$ are
the radial and tangential component of the kinetic energy tensor: a fully
isotropic model have $\xi=1$, while models with $\xi>1.5$ undergo bar
instability on timescales of few tens of dynamical times (Nipoti et al.
2002)\footnote{The generally adopted parametrisation of orbital anisotropy through
the $\beta= 1-({v_t^2}/{v_r^2})$, where $v_t$ and $v_r$ are the tangential 
and radial components of the velocity, respectively, is not particularly
informative for K-M models, since, by construction, they have $\beta=0.0$ at 
their center and $\beta=1.0$ in their outermost region, the shape of the
distribution of $\beta$ being characterised by the anisotropy radius $r_a$ 
\citep[see][and references therein]{iba11}.}.
The maximum degree of radial anisotropy of the considered K-M model
($r_{a}=0.8~r_{h};~\xi=1.5$) is required to obtain a reasonable
reproduction of the extended outer branch of
the SB profile. The velocity dispersion profile provides a good
description of the nearly-flat branch of the observed profile but it fails
to fit the
central point by more than six times the error on the central dispersion
by \citet{dubath97}. It is interesting to note that excluding the central
point to the best-fitting of the velocity normalisation leads to a nearly
perfect fit to the profile, for $r>0.5\arcmin$, but, in this case, the
predicted value of the central dispersion is as high as 14.0~km/s.

%%%%%%%%%%%%%%%%%%%%%%%%%%%%%%%%%%%%%%%%%%%%%%%%%%%%%%%%%%%%%%%%%%%%%%%%
%%%%%%%%%%%%%%%%%%%%%%%%%%%%%%%%%%%%%%%%%%%%%%%%%%%%%%%%%%%%%%%%%%%%%%%%

 In real clusters formed by stars with a mass spectrum and where the
effects of collisions in the cluster center are non negligible, the
density
and velocity dispersion profiles are different for stars of different
mass. Broadly speaking, massive stars tend to transfer kinetic energy to
less
massive stars thus becoming kinematically cooler and sinking in the
innermost region of the cluster on less energetic orbits. The amount of
kinetic
energy transferred by a given star is a function of the contrast between
its mass and the average mass of cluster stars, therefore depending on the
Mass Function (MF).
The three King multi-mass models shown in Fig.~\ref{aniso} as dotted,
dashed and dot-dashed lines have been computed following \citet[][G79 hereafter]{gg79}
assuming 8 mass bins
(all covering equal-mass intervals at different ranges) between 0.1 and
0.8 $M_{\odot}$ populated according to different
assumption on the MF and degree of radial anisotropy:
the dotted lines correspond to an isotropic model with a single power law
MF (in the form $Ndm\propto m^x$), with index $x=-1$, the short-dashed lines
correspond
to an isotropic model with a \citet{kroupa} MF and the long-dashed lines
correspond to a radially anisotropic model ($r_{a}=0.74~r_{h}$) with a
\citet{kroupa} MF. Dark remnants have been added to the original MF
following the prescriptions by Sollima, Bellazzini \& Lee (2012).
Since our SB profile, and especially the velocity dispersion profile, are
mainly based on giant stars we derived the best-fit by comparing them with
the predictions for the most massive mass bin.
While both the isotropic multi-mass models fails to reproduce the outer
branch of the SB profile, they provide a good fit to the dispersion
profile over
the whole radial range. An even better fit is provided by the anisotropic
model (with $r_{a}=0.75~r_{h};~\xi=1.31$) which well reproduce the shape of both the SB and the velocity
dispersion profile along their entire extent.

A gentler decline of the velocity dipersion curve with respect to isotropic K66
models is also predicted in the framework of the MOdified Newtonian Dynamics 
\citep{mond}. However in this case one would expect a convex shape of the velocity dispersion profile
in contrast with our data, unless a high degree of radial anisotropy is 
present \citep[i.e., higher than that assumed here for King-Michie models in Newtonian
dynamics, see][]{sollinip}. In general, given the good fit provided by darkEXP
and multi-mass anisotropic models described above, we conclude that there is no need to invoke a
modification of the Newtonian gravity to explain the observed structure and kinematics of this 
cluster.

%The conclusion of this set of comparisons is that neither a single-mass radially anysotropic K-M model or a multi-mass isotropic model can provide
%a satisfactory fit of both the SB profile and the kinematics of this cluster. On the other hand it seems very plausible that it should exist a
%multi-mass model with some anisotropy that can provide an overall good fit to the data: anisotropy would help to push stars far enough to fit the
%outer branch of the SB profile, energy equipartition would help in lowering the predicted central velocity dispersion.

\subsubsection{Dynamical mass estimates}

It is interesting to note that the masses of the three best-fitting 
isotropic single-mass models are fully consistent. For the K66 model
we obtain $M_{K66}=2.5\times 10^5~M_{\sun}$, for the W75 model
$M_{W75}=2.6\times 10^5~M_{\sun}$,
and for the DARKexp model $M_{HW10}=2.5\times 10^5~M_{\sun}$,
corresponding to $\frac{M}{L_V}\simeq 1.8$.
These values are in excellent agreement with the only previous 
dynamical mass estimate that can be found in the literature,
i.e., $M=2.5\times 10^5~M_{\sun}$ by \citet{PM93}, based on
the central velocity dispersion from a preliminary analysis of the 
data of \citet{dubath97}. The agreement is good also with the
non-dynamical estimates by \citet{mclau05}, who found 
$M=2.1\times 10^5~M_{\sun}$ for both K66 and W75 models, adopting
$M/L_V=1.9$ and a total V luminosity lower than that estimated by C11 and used here
($M_V=-7.8$ instead of $M_V=-8.0$).

On the other hand, anisotropic and multi-mass models suggest a
slightly larger mass, due to the fact that these models have a significant
fraction of their kinetic energy in motions that are not accessible to a sampling
of line-of-sight velocities of giants.
The single-mass K-M anisotropic model gives
$M_{KM}=2.8\times 10^5~M_{\sun}$, the isotropic multi-mass models give
$M_{G79}=3.4\times 10^5~M_{\sun}$ and $M_{G79}=3.6\times 10^5~M_{\sun}$
for
the power-law MF with $x=-1$ and the Kroupa (2002) MF, respectively, and
the anisotropic multi-mass model gives $M_{G79-A}=4.9\times
10^5~M_{\sun}$.
Using the anisotropy-independent estimator of the mass enclosed within the
half light radius by \citet{wolf}, integrating over the observed SB profile 
and interpolating the dispersion profile with a spline,
we obtain $M_{W,1/2}=1.5\times 10^5~M_{\sun}$. Since in a star cluster the
mass should approximately follow light (modulo the mass segregation)
the total mass should be $M_{W,Tot}\sim 3.0\times 10^5~M_{\sun}$,  also
in reasonable agreement with the model-dependent estimates.

\section{Summary and conclusions}
\label{sum}

We obtained RV and [Fe/H] estimates from medium-resolution GIRAFFE spectra for 165 stars selected to lie on the RGB 
of the remote globular cluster NGC~5694. Using both RV and [Fe/H] we selected a sample of 89 bona-fide cluster members, 
83 from the GIRAFFE sample and six from the UVES sample presented in Mu13. Based on these data we derived a mean cluster 
metallicity of [Fe/H]$=-2.01\pm 0.02$, an intrinsic metallicity dispersion of $\sigma_{int}=0.00\pm 0.02$~dex and 
a systemic radial velocity of $V_{sys}=-139.2\pm 0.4$~km/s.

The cluster kinematics is characterised by a very weak systemic rotation, fully consistent with the
{\em rotation - metallicity} and {\em rotation - HB morphology} relations derived by \citet{bellazzini12}. 
The velocity dispersion profile flattens out at large radii. This is incompatible with isotropic single-mass K66 models but is reasonably 
reproduced by both W75 and DARKexp models. However, W75 models provide an unsatisfactory fit to the the cluster SB profile, that, on the other hand is well reproduced by DARKexp models over its whole extent.
While anisotropic single-mass K-M models and multi-mass isotropic King models seem unable to provide an overall good representation of the structure and kinematics of the cluster, we showed that this result can be attained with multi-mass models including anisotropy.

Different models / mass estimators consistently converge on a mass between $M\simeq 2.5\times 10^5~M_{\sun}$, and $M\simeq 4.9\times 10^5~M_{\sun}$ corresponding to $\frac{M}{L_V}\simeq 1.8-3.5$ quite typical for GCs \citep{PM93,solli}. The two models providing the best representation of both the SB and the $\sigma$ profiles lie at the extremes of this range, the DARKexp isotropic non-collisional model at the lower end and the multi-mass anisotropic model (G79-A) at the upper end. The anisotropy-independent mass estimator by \citet{wolf} is in better agreement with the DARKexp model, thus favouring mass (and mass-to-light) values toward the lower end. Low values of the mass-to-light ratio ($M/L_V\simeq 1.8$) are also in agreement with the predictions of population synthesis models, as derived by \citet{mclau05}. We do not report the uncertainties on the individual mass estimates (due to the errors in the input parameters, like, e.g., distance, radii, etc.)
since they are significantly smaller than the the factor of $\sim 2$ systematic uncertainty that is associated to the choice of a given model (DARKexp or G79-A, in particular).

\begin{figure}
\includegraphics[width=84mm]{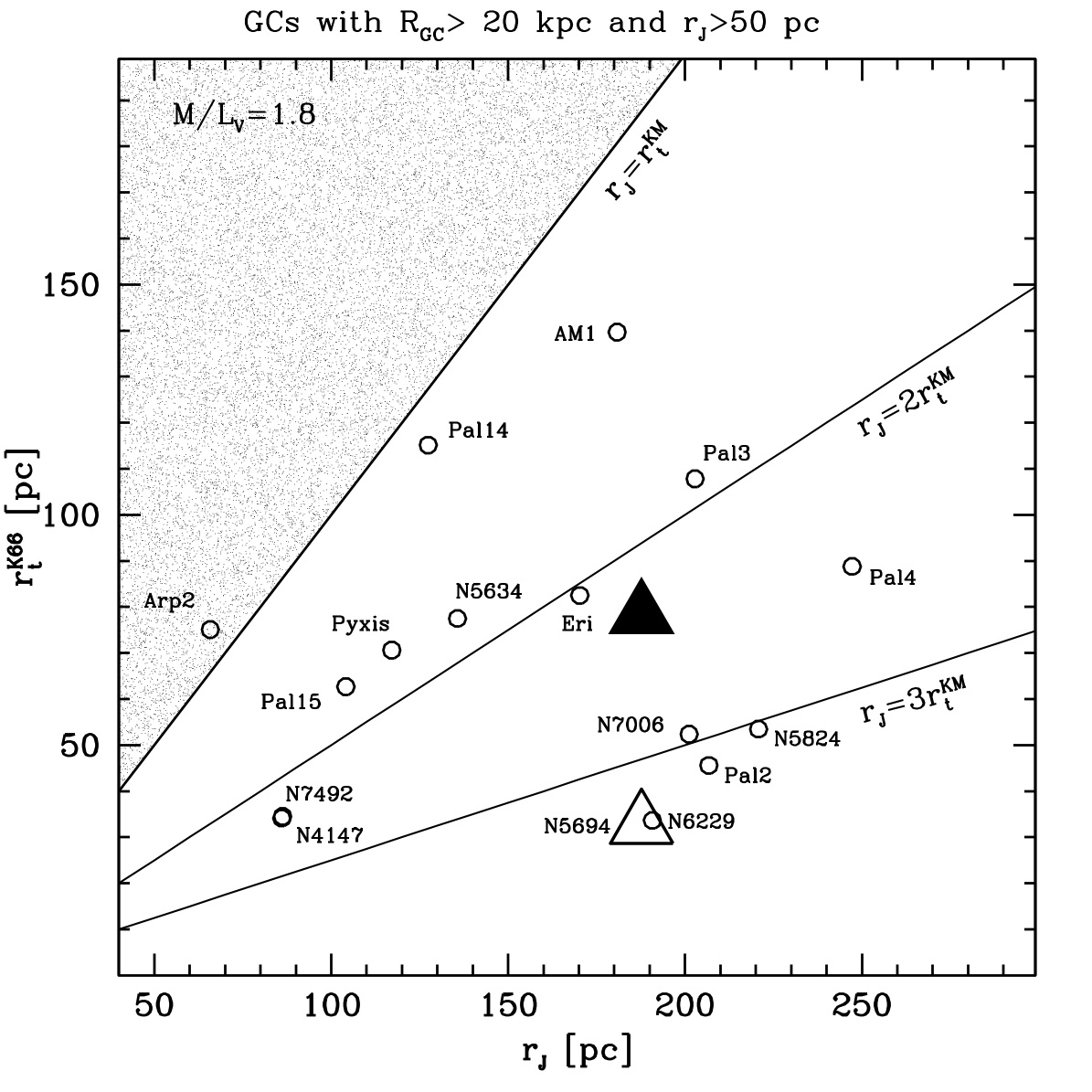}
\caption{Jacoby radii versus tidal radii of the best-fitting K66 model for Galactic GCs more distant 
than 20~kpc from the Galactic  center and having Jacoby radius larger than 50~pc (the latter selection has been introduced 
to make the plot as easy to read as possible; NGC~2419 is not included in the plot since it has a Jacobi radius much 
larger than all the other clusters,$r_J=604.4$~pc). Lines at fixed $r_J/r_t^{KM}$ ratio are also plotted for reference. 
NGC~5694 is plotted as an empty triangle when the K66 tidal radius is adopted, and as a filled triangle when the radius 
of the outermost point of our SB profile is adopted instead. The shaded area correspond to the region of over-filling clusters. }
\label{underf}
\end{figure}

As noted in C11 and clearly illustrated in Fig.~\ref{underf} (produced with the same assumptions as C11, adopting $M/L_V=1.8$, 
and taking tidal radii, total luminosities and Galactocentric distances from \citealt{harris10}) NGC~5694 is largely 
under-filling its Roche lobe, having a ratio between Jacoby radius and tidal radius $r_J/r_t^{KM}>2.0$, independently on the actual assumption on the limiting radius 
(i.e., the tidal radius of the best-fitting K66 model or the outermost point of the observed profile). This holds also 
if the criterion by \citet{baum}, based on the ratio between half-light radius and Jacoby radius ($r_h/r_J<0.05$), is adopted: 
NGC~5694 has $r_h/r_J<0.015$. These authors find that tidally under-filling clusters  form a distinct family (compact clusters) 
with respect to tidally filling ones and concluded that they were likely born compact.
The smooth nature of both the SB and the velocity dispersion profiles of NGC~5694 suggest that indeed it may be tidally 
undisturbed and may represent the typical status of a compact globular cluster evolved in isolation. 
This condition would also favour the permanence of an original radial anisotropic bias 
in the velocity distribution of cluster stars.

In this context, it is interesting to note that (a) virtually all the clusters shown in Fig.~\ref{underf} and 
having $r_J/r_t^{KM}>2.0$ display SB excesses in their outer regions, with respect to K66 models \citep[see, e.g.][]{sohn,JG10}, 
and, in particular (b) many of the brightest among these clusters ($M_V<-7.5$) display smooth power-law profiles incompatible with K66 models in their outskirts, 
similar to NGC~5694 \citep[see, e.g.,][for NGC~7006, NGC~6229, and NGC~5824, respectively]{JG10,sanna12,sanna14}
\footnote{NGC~2419 is well fitted by an anisotropic King-Michie model \citep[see][]{iba11}; Palomar~2 does not appear to have 
an anomalous profile, but the analysis may suffer from the impact of an high (and varying) interstellar extinction \citep{pal2}.}. Deeper and more thorough analyses of the structure and kinematics of these clusters, extending into their low SB outskirts, seems timely. The study of a sample of distant {\em compact} clusters 
\citep[in the sense defined by][]{baum} may provide precious insight on the initial conditions of GCs whose evolution should be only weakly influenced by the interaction with the Milky Way.

\section*{Acknowledgments}

We are grateful to an anonymous referee for useful comments and suggestions.
M.B. and A.S. acknowledges the financial support from PRIN MIUR, project: {\em The
Chemical and Dynamical Evolution of the Milky Way and Local Group Galaxies},
prot. 2010LY5N2T, (PI F. Matteucci). 
Support for M.C., P.A., and C.C. is provided by the Ministry for the Economy, Development, and Tourism's Programa Iniciativa
Cient\'{i}fica Milenio through grant IC\,120009, awarded to the Millennium Institute of Astrophysics (MAS). M.C. also acknowledges
support by Proyecto Basal PFB-06/2007 and by FONDECYT grant \#1141141.

\end{document}